\documentclass[preprint]{aastex}
\usepackage{emulateapj5,apjfonts}
\journalid{}{}
\articleid{}{}
\slugcomment{}

\begin{document}

\def\plotthree#1#2#3{\centering \leavevmode
\epsfxsize=0.30\columnwidth \epsfbox{#1} \hfil
\epsfxsize=0.30\columnwidth \epsfbox{#2} \hfil
\epsfxsize=0.30\columnwidth \epsfbox{#3}}

\title{Optical and Near-Infrared Imaging of the IRAS 1-Jy Sample of 
Ultraluminous Infrared Galaxies. I. The Atlas}

\author{D.-C. Kim\altaffilmark{1,2}, S. Veilleux\altaffilmark{3,5}
, and D. B. Sanders\altaffilmark{4,5}}

\altaffiltext{1}{Institute of Astronomy and Astrophysics, Academia
Sinica, P.O. Box 1-87, Nankang, Taipei, 115 Taiwan}

\altaffiltext{2}{Current address: School of Earth and Environmental
Sciences (BK21), Seoul National University, Seoul, Korea; E-mail:
dckim@astro.snu.ac.kr}

\altaffiltext{3}{Department of Astronomy, University of Maryland,
College Park, MD 20742; E-mail: veilleux@astro.umd.edu}

\altaffiltext{4}{Institute for Astronomy, University of Hawaii, 2680
Woodlawn Drive, Honolulu, HI 96822, and Max-Planck Institut for
Extraterrestrische Physik, D-85740, Garching, Germany; E-mail:
sanders@ifa.hawaii.edu}

\altaffiltext{5}{Visiting Astronomer, W. M. Keck Observatory, jointly
operated by the California Institute of Technology and the University
of California}

\begin{abstract}
An imaging survey of the {\em IRAS} 1-Jy sample of 118 ultraluminous
infrared galaxies was conducted at optical ($R$) and near-infrared
($K^\prime$) wavelengths using the University of Hawaii 2.2m
telescope. The methods of observation and data reduction are
described. An $R$ and $K^\prime$ atlas of the entire sample is
presented along with some of the basic astrometric and photometric
parameters derived from these images. A more detailed analysis of
these data is presented in a companion paper (Veilleux, Kim, \&
Sanders 2002).
\end{abstract}


\section{Introduction}

Ultraluminous infrared galaxies (ULIGs; log [$L_{\rm IR}$/$L_\odot$]
$\ge$ 12 by definition; $H_0$ = 75 km s$^{-1}$ Mpc$^{-1}$ and $q_0$ =
0) are an important class of extragalactic objects that appear to be
powered by a mixture of starburst and AGN activity, both of which are
fueled by an enormous supply of molecular gas that has been funneled
into the nuclear region during the merger of gas-rich spirals. ULIGs
rival quasi-stellar objects (QSOs) in bolometric luminosity, and there
is speculation that ULIGs may indeed represent an important stage in
the formation of QSOs as well as powerful radio galaxies, and that
they may also represent a primary stage in the formation of giant
ellipticals (see Sanders \& Mirabel 1996 for a review).

A complete sample of 118 ULIGs ($f_{60} > 1$) was compiled by Kim
(1995; see also Kim \& Sanders 1998) from a redshift survey of objects
in the {\em IRAS} Faint Source Catalog (FSC: Moshir et al. 1992). As
the nearest (median redshift = 0.145) and brightest ULIGs, this
so-called 1-Jy sample provides the best list of objects for detailed
multiwavelength studies. Optical spectroscopy for 108 objects in the
sample has been published by Veilleux, Kim, \& Sanders (1999a), and
near-infrared spectra have been published for 60\% of the sample by
Veilleux, Sanders, \& Kim (1999b). We now present the results from our
optical and near-infrared ground-based imaging study of this sample.

In this first paper, we discuss the methods of observation and data
reduction (\S 2) and then present an $R$ and $K^\prime$ atlas of the
1-Jy sample (\S 3.1) along with some of the basic astrometric (\S 3.2)
and photometric (\S 3.3) parameters derived from these images. In
Paper II (Veilleux, Kim, \& Sanders 2002, we examine the data in more
detail and combine the results of the image analysis with those from
our spectroscopic survey and other published studies to address the
important issue of the origin of ULIGs and their possible evolutionary
link with QSOs, radio galaxies, and ellipticals.  For both this and
the companion papers, we adopt $H_0$ = 75 km s$^{-1}$, $q_0$ = 0.0,
$M^\ast_R = -21.2$ mag. and $M^\ast_{K^\prime} = -24.1$ mag. The
values for $M^\ast$ are justified in \S 1 of Paper II.

\section{Observations and Data Reduction}

\subsection{Optical Imaging}

Optical images for all 118 ULIGs were obtained at the Cassegrain focus of the
University of Hawaii (UH) 2.2m telescope with the Kron-Cousins $R$
filter (6400 \AA).  These CCD images were bias-subtracted and
flat-fielded using normalized dome flat. Bad columns were removed by
interpolation using nearby pixel values, and cosmic-rays were removed
with the COSMIC package in IRAF.  Except for one object, IRAS
F02021--2103, all objects were observed under photometric conditions
and have been calibrated with the Landolt standard stars (Landolt
1983, 1992).  Atmospheric extinction was corrected using 2--3
observations of the same standard stars at different airmasses through
the night. The IRAF task IMALIGN was used to register the images for a
given object, and the task IMCOMBINE was used to coadd them together.
A summary of our observations is given in Table 1 including object
names, redshifts and infrared luminosities, observing dates, exposure
times, telescopes and instrument used, pixel scales and seeing
estimated from the full widths at half maximum (FWHM) of 
stellar images within the field of view. 

\subsection{Near-Infrared Imaging}

All of the near-infrared images were obtained under photometric
conditions with either the NICMOS3 or QUIRC near-infrared array camera
at the Cassegrain focus of the UH 2.2m telescope using a $K^\prime$
filter.  The central wavelength of the $K^\prime$ filter is
2.1$\mu$m. It is designed to significantly reduce the thermal
background compared to the conventional 2.2$\mu$m K filter (Wainscoat
\& Cowie 1992). The plate scales of the NICMOS3 and QUIRC cameras are
0$\farcs$37 pixel$^{-1}$ and 0$\farcs$1886 pixel$^{-1}$,
respectively.  For observations made prior to April 1993, separate
nearby sky frames were obtained by moving the telescope 90 --
120$\arcsec$ away from the source position, and these sky frames were
then combined to produce final median sky frames.  For the rest of the
observations, sky frames were constructed by median filtering dithered
object frames.  Because the sky brightness changes rapidly, the
median-filtered sky frames were rescaled to the sky values of each
object frame before the sky subtraction.  Prior to this subtraction,
all hot pixels were eliminated from the distribution of pixel values
in dark frames.  The distribution of pixel values is expected to be
Gaussian and if a pixel value lies above or below the 3$\sigma$
values, then we regarded it as hot.  Dome flats were obtained at the
beginning and end of each night by turning a lamp in the dome on and
off. The dark-subtracted flats were obtained by subtracting the
lamp-off flats from the lamp-on flats.  Flat-fielded images were then
coadded after shifting in fractional pixel units with respect to the
object center.  During this procedure we carefully inspected all of
the image frames to see if any hot pixels remained near the target
object.  Any remaining hot pixels near the target were then replaced
with the median value from all other frames at the same position.  The
final coadded images were calibrated from observations of infrared
standard stars (Elias et. al. 1982) using the PHOT or POLYPHOT
packages in IRAF.  Airmass corrections were carried out using the same
procedure as for the optical observations.  The near-infrared
observations are summarized in Table~1.

\section{Empirical Results}

\subsection{The Atlas}

The $R$ and $K^\prime$ greyscale images and contour plots of the
entire 1-Jy sample of ULIGs are presented in Figure 1.  It is
immediately evident from this figure that nearly all of the objects in
the 1-Jy sample show signs of a strong tidal interaction/merger in the
form of distorted or double nuclei, tidal tails, bridges, and
overlapping disks. These results are qualitatively consistent with
previous imaging surveys of ULIGs (e.g., Sanders et al. 1988; Melnick
\& Mirabel 1990; Murphy et al. 1996; Clements et al. 1996; although
see Lawrence et al. 1989; Zou et al. 1991; Leech et al. 1994).  Recent
surveys with HST and adaptive optics systems have also confirmed these
results (e.g., Surace et al. 1998; Zheng et al. 1999; Scoville et
al. 2000; Borne et al. 2000; Cui et al. 2001; Farrah et al. 2001;
Colina et al. 2001; Bushouse et al. 2002; Surace \& Sanders 1999,
2000; Surace, Sanders, \& Evans 2000, 2001).  Great care was used to
determine which objects in the field are involved in tidal
interactions and which ones are not (the description of this analysis
is postponed until \S 3 in Paper II). Faint tidal features and
distortions are generally easier to detect in our $R$-band
images. Galaxies labeled ``G'' in Figure 1 indicates a nearby group
member with no evidence of tidal interaction with the main galaxy. The
group membership of these galaxies was individually determined using a
series of spectroscopic observations with the Keck II telescope. These
observations will be described in Paper II along with the results from
a quantitative analysis of the imaging data.

\subsection {Astrometry}

We have measured the positions of the main galactic nuclei from the
$R$ and $K^\prime$ images.  To determine the position of each nucleus,
we first retrieved the positions of 3 -- 8 stars around the galaxy
from the USNO-A1.0 catalog (Monet et al. 1996).  A plate solution was
then derived using the IRAF task PLTSOL.  The typical positional error
($\sim$ 2-$\sigma$) is estimated to be less than 0$\farcs$5.  The
positions of the nuclei are listed in the second and third columns of
Table 2.  The median values of the positional offsets between the $R$
and $K^\prime$ nuclei is 0$\farcs$48, or 1.22 kpc if we convert
arcseconds into physical scale.  Using the optical spectral types
derived from our optical spectroscopy (Veilleux et al. 1999a), we find
that the median values of the offsets are 0$\farcs$51 (1.3 kpc),
0$\farcs$59 (1.2 kpc), 0$\farcs$44 (1.3 kpc), and 0$\farcs$22 (0.9
kpc) for H~II region-like, LINER, Seyfert 2, and Seyfert 1 galaxies,
respectively. This marginally significant positional offset can be
attributed to the wavelength-dependent effects of dust extinction
along our line of sight. The smaller offset among Seyfert 1s may
indicate less dust extinction in these objects.

\subsection {Aperture photometry}

Global and nuclear (4-kpc diameter) aperture photometry was performed
for each galaxy in the sample. The results of these measurements are
listed in Table 3. Since the 1-Jy sources are selected to have $\vert
b \vert$ $>$ 30$^\circ$ (Kim \& Sanders 1998), the photometry
presented in this paper has not been corrected for Galactic
extinction. Figure 2 shows the distribution of the integrated (=
nuclear + host) $R$ and $K^\prime$ absolute magnitudes of our sample
galaxies.  The absolute magnitudes range from --20.1 to --26.4 at $R$
(mean = --21.9 $\pm$ 0.9, median = --21.83) and from --23.7 to --29.3
at $K^\prime$ (mean = --25.3 $\pm$ 1.1, median = --25.2).  Using $M_R$
= --21.2 and $M_{K^\prime}$ = --24.1 for an $L^\ast$ galaxy with $H_0$
= 75 km s$^{-1}$ Mpc$^{-1}$ (see discussion at the end of \S 1 in
Paper II), these magnitudes are equivalent to $\sim$ 0.4 -- 120 (0.7
-- 120) $L^\ast$ at $R$ ($K^\prime$), with means and medians around 2
(3) $L^\ast$.

Figure 3 shows the distribution of global and nuclear $R - K^\prime$
colors.  The global $R - K^\prime$ and nuclear (4-kpc diameter) $(R -
K^\prime)_4$ colors range from 2.15 to 5.28 and 3.18 to 6.81 with
median values of 3.25 and 4.37, respectively. There is a strong
correlation between the global and nuclear $R - K^\prime$ colors (not
shown here); this correlation is expected since the nuclear color
contributes significantly to the global color.  These colors can be
compared with those of normal galaxies using the photometric tables of
de Vaucouleurs \& Longo (1988). In their table, we find 14 galaxies (5
ellipticals, 9 spirals) which have Kron-Cousins R and near-infrared
$K$ and $H$ magnitudes; for these objects we find a mean $R -
K^\prime$ value of 2.62 $\pm$ 0.34. The $H$ magnitude was used to
convert $K$ to $K^\prime$ magnitudes using the conversion formula
$K^\prime = K + (0.22 \pm 0.03) (H - K)$ from Wainscoat \& Cowie
(1992).  Dust extinction in the nuclei of ULIGs is at least partially
responsible for the redder colors of ULIGs relative to normal
galaxies; dust emission at $K^\prime$ may also be important in
ULIGs. This issue is discussed in more detail in Paper II.

We define the ``compactness'' of a galaxy as the ratio of nuclear
(4-kpc diameter) to total luminosities. Figures 4 shows the histograms
of the $R$ and $K^\prime$ compactness values. The $R$ ($K^\prime$)
compactness ranges from 0.03 (0.10) to 0.53 (0.91) with a mean value
of 0.14$\pm$0.09 (0.36$\pm$0.17) (1 $\sigma$).  The ULIGs in our
sample are therefore significantly more compact at $K^\prime$ than at
$R$, possibly a consequence of the lower extinction and stronger
warm-dust emission at longer wavelengths. The dependence of the
compactness on the infrared and optical spectroscopic properties of
the galaxies will be discussed in Paper II.

\section{Summary}

We have carried out an optical ($R$) and near-infrared ($K^\prime$)
imaging survey of the IRAS 1-Jy sample of 118 ULIGs.  In this first
paper, we present an atlas of the sample galaxies and list a few basic
astrometric and photometric parameters derived from these
images. Nearly all ULIGs show signs of tidal interaction, in agreement
with previous studies. The 1-Jy ULIGs show a broad range of
luminosities with median $R$-band ($K^\prime$-band) integrated
luminosities of about 2 (3) $L^\ast$. Their nuclear and, to a lesser
extent, global $R - K^\prime$ colors are significantly redder than
those of normal galaxies, a result which is attributed to dust
reddening and emission in the inner 4 kpc of these galaxies. This is
also consistent with the fact that the brightness distributions of
1-Jy ULIGs are more compact at $K^\prime$ than at $R$. The generally
smaller nuclear offsets between $R$ and $K^\prime$ images among
Seyfert 1s may indicate lower dust extinction in these objects, but
this result is only marginally significant.  A more detailed analysis
of these data is presented in a companion paper (Veilleux, Kim, \&
Sanders 2002; Paper II).

\clearpage

\acknowledgments

We acknowledge the help from Jim Deane, Aaron Evans, Cathy Ishida, and
Joe Jensen in acquiring some of the images presented in this paper.
We thank the anonymous referee for a prompt and thorough review. S.V.
is grateful for partial support of this research by NASA/LTSA grant
NAG 56547.  D.C.K.  acknowledges financial support from the Academia
Sinica in Taipei and the BK21 project of the Korean
Government. D.B.S. gratefully acknowledges the hospitality of the
Max-Planck Institut for Extraterrestriche Physik and is grateful for
support from a senior award from the Alexander von Humboldt-Stiftung
and from NASA JPL contract 961566. This work has made use of NASA's
Astrophysics Data System Abstract Service and the NASA/IPAC
Extragalactic Database (NED), which is operated by the Jet Propulsion
Laboratory, California Institute of Technology, under contract with
the National Aeoronautics and Space Administration.

\clearpage

\begin{figure}
\caption{$R$ (left) \& $K^\prime$ (right) greyscale images and
contour plots of the 118 objects in the 1-Jy sample of ultraluminous
infrared galaxies.  Each tick mark represents 5$\arcsec$ and the thick
vertical bar along the separating line of each pair of panels
represents a physical scale of 20 kpc.  The bottom contour levels
(BCLs) of the $R$ \& $K^\prime$ images are given in the last two
columns of Table 2, and the contour intervals are separated by 0.5
magnitude.  Objects labeled ``G'' in the figure indicate nearby group
members with no evidence of tidal interaction with the main system,
and the ``*'' symbol indicates a star.  The other labels (N, S, E, W,
etc) identify the nuclei of parent galaxies and are so labeled in
Tables 2 and 3.}
\end{figure}

\begin{figure}
\caption{$R$ and $K^\prime$ absolute magnitudes. These galaxies span a
broad range of total (= nuclear + host) luminosities. For comparison,
$M_R$ = --21.2 and $M_{K^\prime}$ = --24.1 for an $L^\ast$ galaxy.}
\end{figure}

\begin{figure}
\caption{Global and nuclear (4-kpc diameter) $R - K^\prime$
colors. These colors (especially the nuclear values) are considerably
redder than those of normal galaxies, $<R - K^\prime> = 2.62 \pm 0.34$
(1 $\sigma$). Dust is the suspected culprit.}
\end{figure}

\begin{figure}
\caption{$R$ and $K^\prime$ compactness values. This quantity is defined as
the ratio of the nuclear (4-kpc diameter) to total luminosity. ULIGs
are more compact at $K^\prime$ than at $R$, another sign that dust
affects the central surface brightnesses of these objects.}
\end{figure}

\clearpage
\setcounter{figure}{0}

\begin{figure}
\epsscale{0.85}
\caption{}
\end{figure}

\clearpage

\begin{figure}
\epsscale{1.25}
\plotone{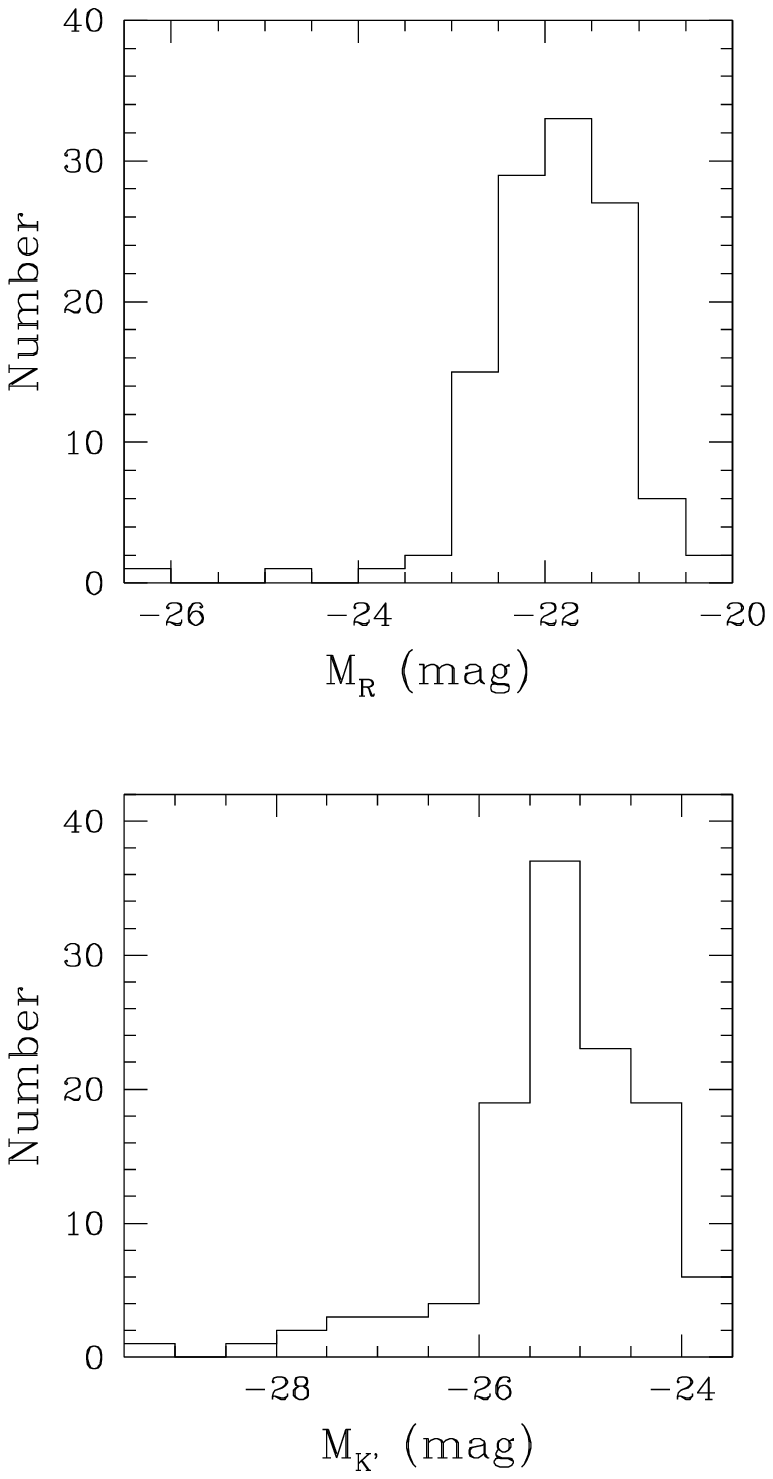}
\caption{}
\end{figure}

\begin{figure}
\plotone{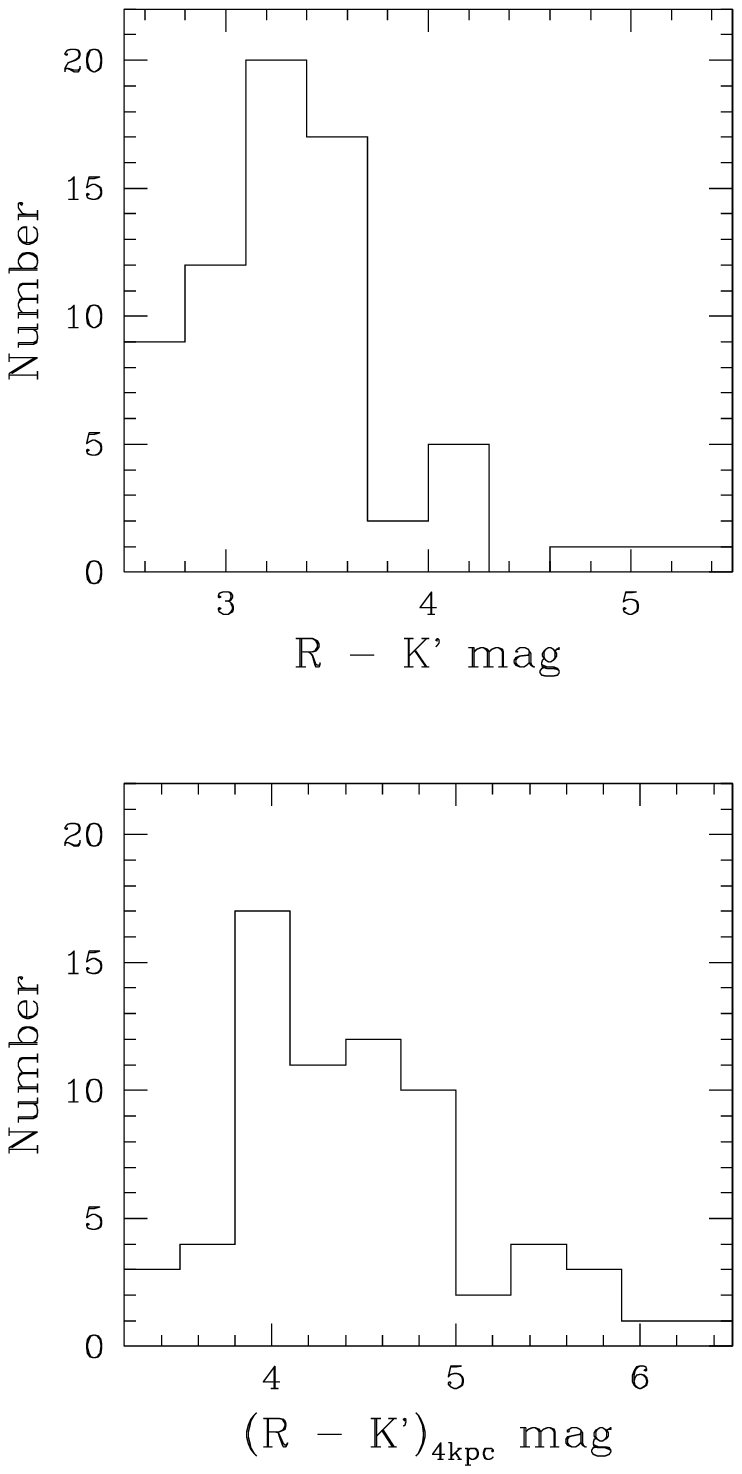}
\caption{}
\end{figure}

\begin{figure}
\plotone{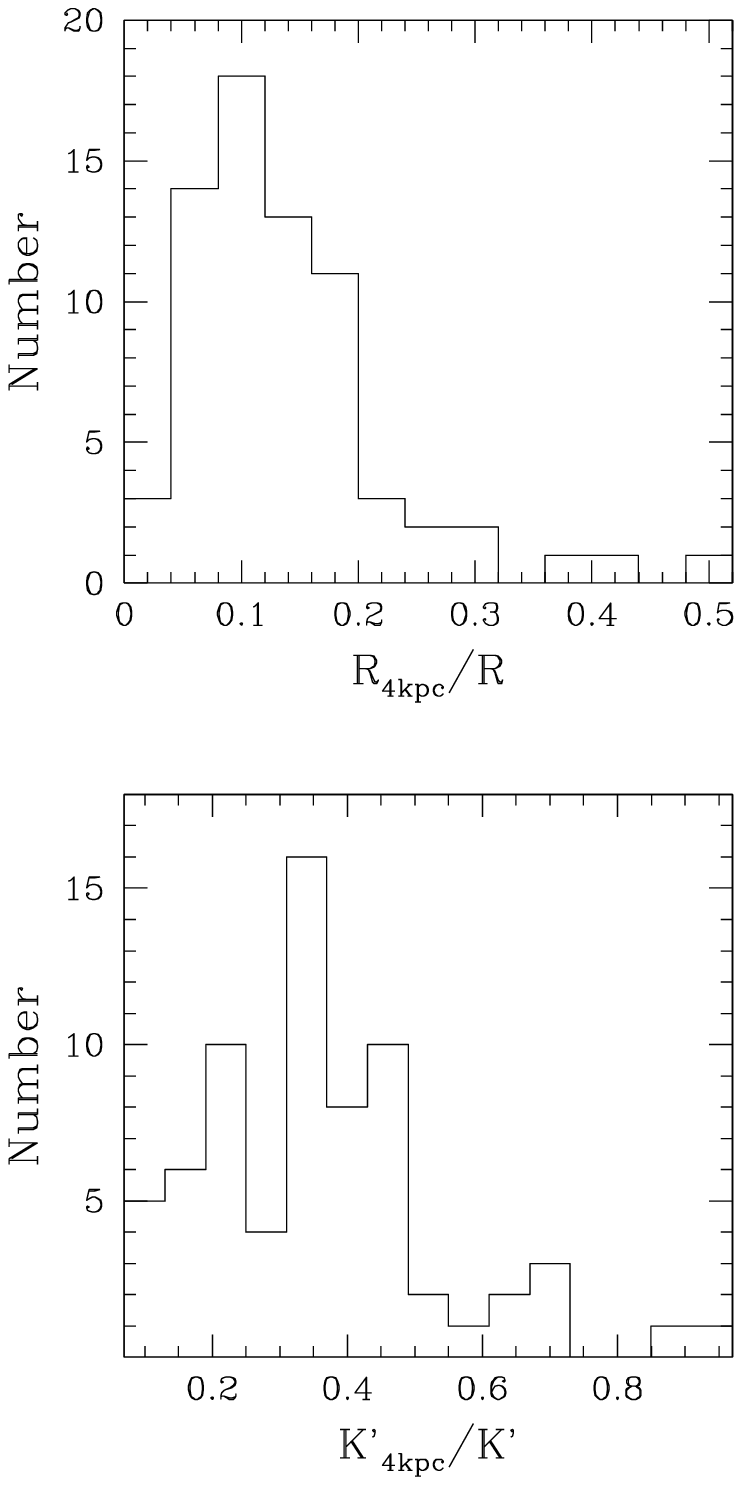}
\caption{}
\end{figure}

\clearpage

\end{document}


\newcommand{\ts}{\thinspace} 
